\documentclass[fleqn,usenatbib]{mnras}

\usepackage{newtxtext,newtxmath}

\usepackage[T1]{fontenc}
\usepackage{float}

\DeclareRobustCommand{\VAN}[3]{#2}
\let\VANthebibliography\thebibliography
\def\thebibliography{\DeclareRobustCommand{\VAN}[3]{##3}\VANthebibliography}

\usepackage{graphicx}	
\usepackage{amsmath}	

\def\be{\begin{equation}}
\def\ee{\end{equation}}
\def\beq{\begin{eqnarray}}
\def\eeq{\end{eqnarray}}
\def\gtrsim{\mathrel{\hbox{\rlap{\hbox{\lower4pt\hbox{$\sim$}}}\hbox{$>$}}}}
\def\lsrsim{\mathrel{\hbox{\rlap{\hbox{\lower4pt\hbox{$\sim$}}}\hbox{$<$}}}}

\def\3half{\frac{3}{2}}

\title[H, He-like recombination spectra VI]{H, He-like recombination spectra VI: Quadrupole $l$-changing collisions}

\author[E. Deliporanidou et. al.]{
E. Deliporanidou,$^{1,2}$\thanks{E-mail: ed650@cam.ac.uk}
N. R. Badnell,$^{1}$  P. J. Storey,$^{3}$ G. Del Zanna,$^{2,4}$ G. J. Ferland$^{5}$
\\
$^{1}$Department of Physics, University of Strathclyde, Glasgow G4 0NG, UK\\
$^{2}$DAMTP, Centre for Mathematical Sciences, University of Cambridge, Wilberforce Road, Cambridge, CB3 0WA, UK\\
$^{3}$Department of Physics and Astronomy, University College London, London WC1E 6BT, UK\\
$^{4}$School of Physics \& Astronomy, University of Leicester, Leicester  LE1 7RH, UK \\
$^5$Department of Physics and Astronomy, University of Kentucky, Lexington, KY 40506, USA}

\date{Accepted XXX. Received YYY; in original form ZZZ}

\pubyear{\the\year{}}

\begin{document}
\label{firstpage}
\pagerange{\pageref{firstpage}--\pageref{lastpage}}
\maketitle

\begin{abstract}
We have developed a simple analytic formula that well describes quadrupole $l$-changing collisions of the form $nl \rightarrow nl'$, as confirmed by comparison with numerical quantal Born calculations obtained with the program {\sc autostructure} (Badnell 2011). 
Such formulae could easily be included in models of astrophysical
plasma emission, such as the hydrogen and helium-like recombination spectra. 
When compared with the results of previous quantal calculations based upon an analytic solution of the time-dependent Schr\"odinger equation by Vrinceanu \& Flannery (2001), we find relatively good agreement, 
with the exception of large $l > n/2$ transitions. 
We provide a tentative explanation for such discrepancies. 
However, we also show that the rates for quadrupole $l$-changing collisions 
are typically two orders of magnitude lower than the dipolar ones. Inclusion of the 
quadrupolar rates in a hydrogenic collisional-radiative model of nebular plasma shows minimal changes 
to the level populations, typically within 1\% in nebular conditions. Simple and complete theories are now available for $l$-changing collisions suitable for astrophysical applications. 
\end{abstract}

\begin{keywords}
atomic data -- ISM: abundances -- ISM: H~{\sc ii} regions --
cosmology: observations --  primordial nucleosynthesis
-- ISM: radio lines
\end{keywords}



\section{Introduction}

Astrophysical plasma, comprising 95\% of cosmic baryonic matter, consists of 
a mixture of ions,  atoms, and free electrons and protons.
Given the large abundance of hydrogen and helium, accurate 
non-LTE collisional-radiative modelling including several ionization 
and recombination processes is required for a vast amount of 
astrophysical applications. 
Long ago, it was recognised that dipole $l$-changing collisions
by protons and electrons are fundamentally important for such 
models \citep{Pengelly1964}. 
$l$-changing collisions are also generally important for redistributing
the populations of atomic states of any atom. 
For this reason, there has been recent interest in improving 
upon the seminal work of \cite{Pengelly1964}, see
 our previous paper \citep{Badnell2021} and references therein. 

This paper focuses on quadrupole $l$-changing collisions for hydrogen,
addressing the need for an investigation that is presently absent in the literature.
Such collisions are 
of the form $n l \rightarrow n l'$, where |$l -l'$| = $\Delta l =2$ and for which we can neglect the excitation energy ($\Delta E \approx 0$). In paper II of this series by \cite{Guzman17}, the effects of higher multipole $l$-changing collisions were briefly examined using the Simplified Semi-Classical (SSC) method of \cite{Vrinceanu2012} for He-like ions, see section~\ref{SSC section}. It was found that while multipole effects are generally negligible, they can contribute up to 50 \% of the differences between the Quantum Mechanical (QM), see section~\ref{QM TDSE section}, and the \cite{Pengelly1964} method (PSM) results, for certain Brackett and Paschen lines at common astrophysical densities. Notably, these differences in line intensities do not exceed 2\%. In this paper we extend this investigation to quadrupole $l$-changing collisions for hydrogen, by providing an analysis of their impact. As in paper II, we explore the role of different approximations and examine the significance of quadrupole transitions under relevant conditions.

The Time-Dependent Schröndiger Equation (TDSE) Quantum Mechanical (QM) approximation and the Born approximation are the primary tools that can be used to solve quadrupole $l$-changing collisions. 
An initial comparison between the two reveals a discrepancy, particularly 
at large $nl \rightarrow nl'$ transitions. Consequently, a deeper analytical comparison was undertaken, employing two additional rate coefficient expressions representing accurate TDSE QM and Born approximations to search for the root causes of the observed disagreement. The Born approximation is a well-known QM approximation for higher multipole expansions. The Simplified Semi-Classical (SSC) approximation developed by \cite{Vrinceanu2012} and a rate coefficient expression developed by \cite{Burgess2005} constitute the TDSE QM and the Born approximations, respectively. Both expressions necessitate simplifications and modifications to account for quadrupole $l$-changing collisions effectively.

In section~\ref{sec:theo}, we discuss the Born and QM rate coefficients. In section~\ref{methodoloy}, we discuss the methods we used for the study. In section~\ref{sec:results}, we present the outcomes of the numerical evaluation and analytic comparison, identifying agreements and discrepancies between the approximations. In section~\ref{Applications}, we discuss the applications of our analysis by testing the relevance of quadrupole rates. In section~\ref{sec:concludes}, we elaborate on the outcomes of the numerical analysis. 

\section{Theory}
\label{sec:theo}

The key quantity used for collisional radiative modelling is the 
rate coefficient [cm$^3$ s$^{-1}$]
\citep[see, e.g.][]{Burgess2005} 
\begin{equation}
    q_{i\rightarrow{}j} = 
    \int f(v) \, v \, Q_{ij} \, \mathrm{d}v\, ,
    \label{BT first RC}
\end{equation} 
where $f(\mathrm{v})$ is the  velocity distribution of the free particles,
assumed to be Maxwellian (a generally good approximation for astrophysical plasma), and $Q_{ij}$ is the 
cross-section for the $l$-changing collision between the target 
states $i$ and $j$. 

The high-energy scattering regime of atoms is characterized by the ratio \( E / \Delta E \), where \( E \) represents the thermal energy and \( \Delta E \) is the excitation energy. This regime requires a large ratio, meaning the thermal energy significantly exceeds the excitation energy. Quadrupole \(l\)-changing collisions exhibit an energy dependence when \( \Delta E \) is finite. When \( E / \Delta E \) becomes large, the regime approaches an infinite energy limit, effectively making the cross-section independent of scattering energy. This limit, in which \( \Delta E \rightarrow 0 \), was detailed by \cite{Burgess2005} in their study of low-\(l\) proton-electron collisions involving fine-structure excitations with zero excitation energy.

The electric quadrupole radiation is much weaker than the electric dipole radiation. In quadrupole proton-electron collisions within a hydrogen atom, the proton changes the electron's angular momentum by 2 units due to angular momentum exchange. Without a cut-off parameter, the cross-section for dipole $l$-changing collisions would diverge. However, quadrupole cross-sections do not diverge, so no cut-off is necessary. 

The interactions asymptotically vary as $1/r^2$ for the dipole, and as $1/r^3$ for quadrupole $l$-changing collisions.

\subsection{The Born approximation}

The Born approximation treats the collider as a plane wave.The total cross section is obtained by summing over a multipole expansion. For scattering between atoms and electrons, the Born cross-section as coded by NRB within 
 {\sc autostructure } \citep{Badnell2011_AUTOSTRUCTURE} is widely used. 
The calculations follow the descriptions given for the different types 
of transitions by \cite{Burgess1997} and \cite{chidichimo_etal:2003},
which refer to the original works by  \cite{born_quantenmechanik_1926} and \cite{Bethe_1930}.

The Born approximation is well-suited to describing proton--ion collisions, especially quadrupole  $l$-changing collisions.
We present the full Born approximation for quadrupole $l$-changing collisions in appendix~\ref{append_born}. This was derived by NRB based on the above
papers, using further assumptions and approximations to make it suitable for quadrupole $l$-changing collisions. The final Born cross-section expression is given by

\be
 Q_{nl \rightarrow nl'}^{\rm Born} = \frac{\pi}{K_q^2 (2l+1)} \times 8 \int \, \frac{\mathrm{d}K}{K^3}\ 4\pi \sum_{\lambda} |<t' \parallel \mathbf{B_{\lambda}} \parallel t>|^2 , 
\label{Borncross}
\ee

where the collision strength is the integral over the momentum transfer, given by 

\be
\Omega_{nl \rightarrow nl'}^{\rm Born} = 8 \int \, \frac{dK}{K^3}\ 4\pi \sum_{\lambda} |<nl' \parallel \mathbf{B_{\lambda}} \parallel nl>|^2 , 
\label{Born Collision strength}
\ee

and, $|<nl' \parallel \mathbf{B_{\lambda}} \parallel nl>|^2$ is the  term in  the summation for multipole 

\begin{equation}
\begin{split}
<n'l' \parallel \mathbf{B_{\lambda}} \parallel nl> \, = (-1)^{l'} \, \left[\frac{(2l+1)(2\lambda+1)(2l'+1)}{4\pi}\right]^{1/2} \\
\times \begin{pmatrix}
l' & \lambda & l\\
0 & 0 & 0
\end{pmatrix} \, <l' \parallel \mathbf{j_{\lambda}} \parallel l> , 
\end{split}
\label{matrix element 1}
\end{equation}

where the 3-j symbol is equal to $0$, unless the sum of the terms is even and $|l-l'| \le \lambda \le (l+l')$, based on the triangular rule of the Wigner 3-j symbol that determines the allowed multipoles of the Born approximation. $K_q$ is the initial momentum of the free electron, K is the momentum transfer and $\lambda$ indicates the multipole. 

The 3-j symbol also mirrors the conservation of parity during interactions, influencing the progression of alternating multipole sequences. In cases of parity conservation, transitions between angular momenta $l$ and $l'$ yield electric multipole series in the form $1-3-5-7$ or $2-4-6-8$, determined by the parity of the electric multipole operator, represented as $(-1)^\lambda$.

The radial part of the cross-section integral involves a spherical Bessel function as presented below, 

\begin{equation}
    <l' \parallel \mathbf{j_{\lambda}} \parallel l> = \, \int_{0}^{\infty} \, P_{n' l'}(r) \, [j_{\lambda}(K r) - \delta_{\lambda 0}] \, P_{nl}(r) \, \mathrm{d}r\ .
    \label{matrix element 2}
\end{equation}

There is no analytic expression for the integral of the spherical Bessel function between two wave functions, even for hydrogen. \cite{Burgess2005} developed simple expressions for proton-electron collisions, which are similar to Born for $l$-changing collisions, as the excitation energy is zero and the collision strength is a constant.

\subsection{Burgess and Tully Modified Born (BTM)}
\cite{Burgess2005} gave an analytic approximation to the full Born cross section for the
general case $(\Delta E \ne 0)$ which we denote BT,
but it depended on a numerical quadrupole line strength ($S_2$) and an effective target radius ($r_0$),
the latter to be determined by matching to numerical Born results. Their expression was initially provided for modelling proton collisions. They obtained cross sections, collision strengths, and rate coefficients with accurate behaviour at infinite energies and higher temperatures, correcting earlier mistakes. 
As we wanted to obtain an analytic expression for our comparisons, 
we have specialised their approach to the $\Delta E =0$ case. We denote 
this as the modified Burgess-Tully (BTM) method.

The BTM rate coefficient, $q^{\rm BTM}$, is related to the collision strength $\Omega^{\rm BTM}$ via

\be
q_{nl\rightarrow nl'}^{\rm BTM}  =\frac{2}{\omega_{l}}
\left(\frac{\pi \, I_H}{kT}\right)^{1/2}
    \mu^{-3/2} \; \Omega^{\rm BTM}_{nl- nl'} \; \exp{(-E/kT)} \; \frac{a_0^3}{\tau_0}\,,
\label{BTMrate}
\ee

where $\omega_{l}= 2l+1$, is the statistical weight , $\mu = M/m_e$ is the dimensionless reduced mass, $\sqrt{M}=30.31$ and $k=6.33\times10^{-6}$ [Ryd/K] in atomic units. $E$ is the excitation energy which is zero, and hence, $\exp{(-E/kT)}=1$. \cite{Burgess2005} introduce a quantity $Y_{ij}$ (eq. 35) in their rate coefficient expression (eq. 34), however, the quantity $Y_{\rm ij}$ is equal to the collision strength $\Omega_{\rm ij}$, because $\Omega^{\rm BTM}$ is independent of energy.

The quadrupole Born collision strength in its high-energy form is given by

\be
\Omega_{nl- nl'}^{\rm BTM}=\frac{4}{5}\left(\frac{\mu Z_p}{r_0}\right)^2 S_2(nl- nl')\,,
\label{collstr}
\ee

where $Z_p$ is the charge number of the projectile,
$S_2$ is the quadrupole line strength, and $r_0$ is the
effective target radius \citep{Burgess2005}. The BT expression used the Bethe approximation and quadrupole line strength from the literature. Their work is focused on proton excitation of the fine structure of the ground term in complex ions, as opposed to the $l$-changing collisions in Rydberg hydrogen, which are effectively degenerate. The high energy limit of the BT formalism corresponds to the $\Delta E = 0$ modification in the BTM expression.

An analytic form for $S_2$ does not appear to be readily available in the literature.
We have used the ladder operator techniques of \cite{Hey2006} to obtain one:

\be
S_2=\left( \frac{300}{2} \right) \left( \frac{n}{2 Z_t} \right)^4 \left (\frac{l_>(l_>-1)}{(2l_>-1)} \right)
      (n^2-l_>^{2})(n^2-(l_>-1)^2) , 
\label{line2}
\ee 
where $Z_t$ is the charge number of the target, and 
$l_>= {\rm max}(l,\allowbreak l')$
(see details in appendix~\ref{append_BTM}).
We have verified this expression against the numerical results from {\sc autostructure}. 

For the effective target radius $r_0$, we adopt an expression of the form 

\be
r_0 = \frac{c_0 \, n^2}{Z_t} , 
\label{r_0 equation}
\ee

obtained based on the turning points of the asymptotic behaviour of the bounded and unbounded solutions of the Schrödinger's radial equation, as  $r \rightarrow \infty $. 

We find the best agreement between our
analytic approximation and exact Born collision strengths if we take

\be
c_0=1+\left[1-\frac{l(l+1)}{n^2}\right]^{1/4}\, .
\label{turnout}
\ee

If the power in equation \eqref{turnout} were $1/2$, then $r_0$ would be the outer turning point of the
(hydrogenic) radial function.

\subsection{TDSE}
\label{QM TDSE section}

\cite{Vrinceanu2001} solved the TDSE for a colliding particle at large impact parameters by making use of the Stark effect in Rydberg transitions. They investigated collisions of slow charged particles at large impact parameters, and they provided an expression for the Stark mixing-transitions’ probability of arbitrary angular momentum Rydberg atomic states of the form $nl \rightarrow nl'$. This method has been evaluated for dipole $l$-changing collisions, but its accuracy for higher multipole collisions hasn't been closely reviewed yet. 
The TDSE quantum mechanical probability

is given by \cite{Vrinceanu2001}

\beq
\label{VR01 QM Prob}
    P_{nl \rightarrow nl'}^{\rm QM} &=& (2l'+1) \sum_{L=|l'-l|}^{n-1} (2L+1) 
\left\{\begin{array}{ccc} l' \quad l \quad L \\ j \quad j \quad j \end{array}\right\}^2
\\ \nonumber
&\times &
\frac{(L!)^2 (n-L-1)!}{(n+L)!} \, (2 \sin \chi)^{2L} \, \left[C_{n-L-1}^{(L+1)} (\cos \chi)\right]^2\,,
\eeq
where $\{...\}$ denotes a Wigner $6j$-symbol, $C_n^{(l)}$ is an ultra-spherical polynomial,
and $j=(n-1)/2$. This is evaluated analytically using techniques from \cite{edmonds1996angular}. The pseudo-multipole summation over $L$ can contribute significantly
in the quadrupole case, unlike the dipole case where the $L=1$ term dominates. The rotation angle $\chi$ is written in terms of the scattering parameter $\alpha$ via

\be
   \cos \chi = \frac{1+\alpha^2\cos(\Delta\Phi\sqrt{1+\alpha^2})}{1+\alpha^2} \quad\quad
   \mbox{and}\quad\quad \alpha =  \frac{3Z_p n \hbar}{2m_e vb}\,,
\label{VR01 cosχ}
\ee

where $\Delta \Phi =\pi$, $b$ is the impact parameter and $v$ is the projectile speed.
Integration over $\alpha$ yields both cross sections and rate coefficients.

\cite{Vrinceanu2012} further evaluated the TDSE QM probability term, and they re-expressed the cross-section,  considering the probability dependence on the impact parameter b and the projectile velocity v, through $\alpha$, which is given by 

\be
 Q_{nl \rightarrow nl'}^{\rm TDSE} = \frac{9\pi}{2} \times \left(\frac{Z_p n \hbar}{m_e \rm v} \right)^2 \times  I_{l \rightarrow l'}^{(n)} \; , 
\label{Vrinceanu Cross Section 2}
\ee

where $I_{l \rightarrow l'}^{(n)}$ is the velocity-independent integral factor that \cite{Vrinceanu2012} introduced, which is determined by the initial and final states as 

\begin{equation}
    I_{l \rightarrow l'}^{(n)} = \int_{0}^{\infty} P_{nl \rightarrow nl'}^{(n)} (\alpha) \,\frac{\mathrm{d}\alpha}{\alpha^3} \ ,
    \label{VR eq.5}
\end{equation}

where the $1/\alpha^3$ dependence generates the logarithmic singularity in the cross-section for large impact parameters. It is also where the cut-off parameter is applied for dipole $l$-changing collisions.

The upper limit $L$ restricts the contribution of higher multipoles in the summation of eq. \eqref{VR01 QM Prob}, which ultimately yields the cross section and rate coefficient. Adjusting this limit controls the impact of higher multipoles on the numerical rate coefficient. For $l \lnapprox n/2$ the 6-j symbol truncates the sum at $L = l+l'$, however, for $l \gnapprox n/2$, the upper limit $L=n-1$ of the summation is the term that truncates the sum.

\subsection{SSC}
\label{SSC section}
The Simplified Semi-Classical (SSC) approximation is an analytic simplification of the QM approximation, without making use of the Bethe approximation and hence no cut-off parameter is introduced for the SSC. This approximation is very accurate at low impact parameters and was derived by \cite{Vrinceanu2012}:

\beq
\label{SSCeq_1}
    q_{nl\rightarrow{}nl'}^{\rm SSC} &=& 1.294 \times 10^{-5} \sqrt{\frac{M}{m_{e}}} \frac{Z_p^2}{\sqrt{T}} 
\\ \nonumber
&\times &
    \frac{n^2[n^2(l+l')-l_{<}^2(l+l'+2|\Delta l|)]}{(l+1/2)|\Delta l|^3} \quad [\rm cm^3 \rm s^{-1}]\, ,
\eeq

where, $l_<$ is the smaller of $l$ and $l'$, $\sqrt{M}=30.3039$, $m_e=1$ in atomic units, and $Z_p=1$. 

The SSC approximation shows high accuracy on quadrupole and higher multipole $l$-changing collisions, compared with the old Semi-Classical (SC) \citep{Vrinceanu2001}. 

\section{Methodology}
\label{methodoloy}

We employed both numerical and analytical methods to compare cross section and rate coefficient expressions. Numerical results for the Born collision strengths are derived from {\sc autostructure}. In addition, QM, SSC and BTM cross sections and rate coefficients are derived by a 
FORTRAN executable program written by NRB. The  program calculates Maxwellian rate coefficients and cross sections for approximations that describe $l$-changing atomic collisions. The Born rate coefficient expression is determined analytically, using the BT rate coefficient and cross section results from the modified $l$-changing collisions program, and the Born collision strength results from the {\sc autostructure} code, where we account for proton collisions by scaling the outputs with the $\sqrt{m_p/m_e}$ ratio.

Expressions displaying discrepancies for identical values of $n$ and $l$ are subjected to analytical comparison to identify the underlying sources of disagreement. We also investigate the contribution of higher multipoles on the summation terms of the QM and the Born approximations by restricting the contribution of higher multipoles on the Born summation and by adding an upper $L$ limit on the QM summation and via a modification to the {\sc autostructure} program. 

\section{Results}
\label{sec:results}

\begin{figure}
    \centering
    \includegraphics[width=\columnwidth]{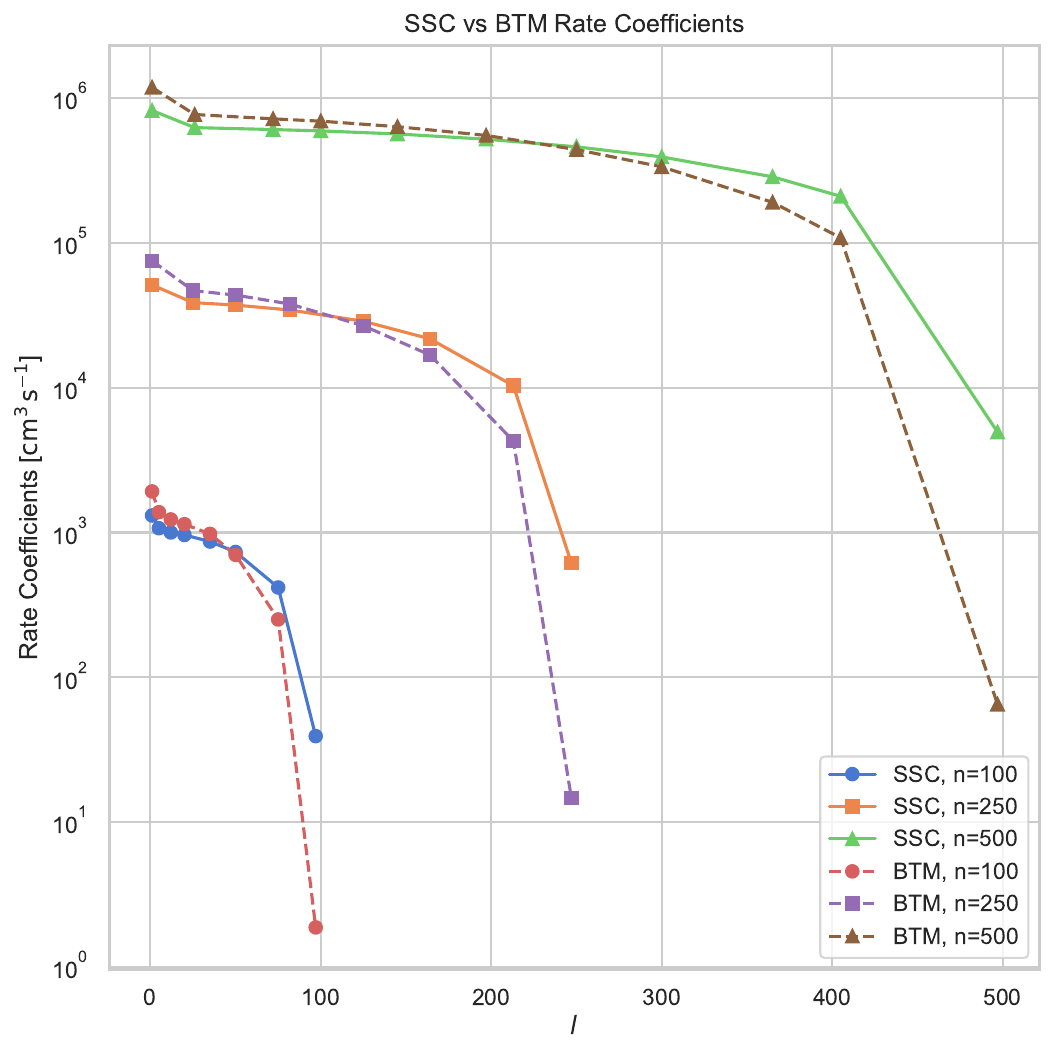}
    \caption{Variation of TDSE QM (SSC) and Born (BTM) rate coefficients [cm$^3$ s$^{-1}$] with $l$ on a logarithmic scale for T=100 [K].}
    \label{fig:SSCvsBTM}
\end{figure}

Our numerical results reveal good agreement between rate coefficients in the range $100<n<500$ across all approximations for low-$l$, with significant discrepancies emerging at higher $l$ values (see fig.~\ref{fig:SSCvsBTM} and table~\ref{table:RC}.

Although the Born is the most accurate approximation for the high energy behaviour, it disagrees with the TDSE QM approximation for intermediate and high $l$ quadrupole $nl \rightarrow nl '$ transitions. Numerical results reveal a good agreement between the BTM and the Born approximations for $c_0=2$.

\begin{table*} \footnotesize
\begin{center}
\caption{
Representative TDSE QM, SCC, BTM, and Born rate coefficients [cm$^3$ s$^{-1}$] at T=100 [K], for various $n$ and $l$ values. All of the higher ($ > 2$) multipoles have been included in the Quantal Born and the TDSE results.
\label{table:RC} 
}
\begin{tabular}{  c  c  c  c  c  c } 
 \hline
 $n$ & $l \rightarrow l' $ & TDSE QM  & SSC  & BTM  & Born  \\ [0.5ex] 
 \hline 
 $20$ & $1-3$ & $3.38 \times 10^0$ & $2.08 \times 10^0$ & $3.04 \times 10^0$ & $2.97 \times 10^0$ \\
 $20$ & $9-11$ & $1.69 \times 10^0$ & $1.25 \times 10^0$ & $1.20 \times 10^0$ & $1.18 \times 10^0$ \\
 $20$ & $17-19$ & $3.34 \times 10^{-1}$ & $3.18 \times 10^{-1}$ & $5.24 \times 10^{-2}$ & $5.52 \times 10^{-2}$ \\
 $30$ & $1-3$ & $1.72 \times 10^1$ & $1.06 \times 10^1$ & $1.57 \times 10^1$ & $1.54 \times 10^1$ \\
 $30$ & $14-16$ & $8.33 \times 10^0$ & $6.19 \times 10^0$ & $6.07 \times 10^0$ & $5.89 \times 10^0$ \\
 $30$ & $27-29$ & $1.10 \times 10^0$ & $1.07 \times 10^0$ & $1.30 \times 10^{-1}$ & $1.34 \times 10^{-1}$ \\
 $50$ & $1-3$ & $1.33 \times 10^2$ & $8.17 \times 10^1$ & $1.22 \times 10^2$ & $1.19 \times 10^2$\\
 $50$ & $24-26$ & $6.30 \times 10^1$ & $4.70 \times 10^1$ & $4.67 \times 10^1$ & $4.52 \times 10^1$ \\ 
 $50$ & $47-49$ & $5.02 \times 10^0$ & $4.93 \times 10^0$ & $4.00 \times 10^{-1}$ & $4.10 \times 10^{-1}$ \\
  $100$ & $1-3$ & $2.13 \times 10^3 $ & $1.31 \times 10^3 $ & $1.96 \times 10^3 $ & $1.92 \times 10^3$ \\ 
  $100$ & $20-22$ & $1.31 \times 10^3$ & $9.61 \times 10^2$ & $1.18 \times 10^3$ & $1.14 \times 10^3$\\
  $100$ & $50-52$ & $9.80 \times 10^2$ & $7.33 \times 10^2$ & $7.25 \times 10^2$ & $7.01 \times 10^2$ \\
  $100$ & $97-99$ & $3.97 \times 10^1 $ & $3.93 \times 10^1$ & $1.80 \times 10^{0}$ & $1.88 \times 10^0$ \\
  $250$ & $1-3$ & $8.34 \times 10^4$ & $5.11 \times 10^4$ & $7.66 \times 10^4$ & $7.50 \times 10^4$ \\
  $250$ & $50-52$ & $4.99 \times 10^4$ & $3.71 \times 10^4$ & $4.52 \times 10^4$ & $4.35\times10^4$ \\
  $250$ & $125-127$ & $3.83 \times 10^4$ & $2.87 \times 10^4$ & $2.87\times 10^4$ & $2.67\times10^4$\\
  $250$ & $247-249$ & $6.15 \times 10^2$ & $6.14 \times 10^2$ & $1.29 \times 10^1$ & $1.47 \times 10^1$ \\
  $500$ & $1-3$ & $1.33 \times 10^6$ & $8.17 \times 10^5$ & $1.23 \times 10^6$ & $1.18 \times 10^6$\\
  $500$ & $100-102$ & $7.91\times10^5$ & $5.91\times10^5$ & $7.19\times10^5$ & $6.90\times10^5$\\  
  $500$ & $250-252$ & $6.13\times10^5$ & $4.59\times10^5$ & $4.61\times10^5$ & $4.39\times10^5$\\
  $500$ & $497-499$ & $4.98 \times 10^3$ & $4.93 \times 10^3$ & $5.66 \times 10^1$ & $6.52 \times 10^1$\\
 \hline
\end{tabular}
\end{center}
\end{table*}

We see from table~\ref{table:RC} that the SSC and BTM approximations are representative of the TDSE QM and Born results respectively over all-$l$. The latter two are also in good agreement with each other at low-$l$. However, for $l \gtrsim n/2$ we see increasing divergence between the TDSE QM/SSC and Born/BTM results. Figure~\ref{fig:SSCvsBTM} shows the variation of the SSC and the BTM with $l$. Similarly with table~\ref{table:RC}, there is agreement between the SSC and the BTM for small $l$ transitions ($l<<n$), better agreement for intermediate transitions ($l \approx n/2$) and large disagreement for $l \approx n$.

Taking the limit of intermediate-$l$-changing collisions ($1<<l<<n$) we find analytically a disagreement of approximately $0.8$ when taking the ratio of $q^{\rm SSC}/q^{\rm BTM}$:

\begin{equation}
    q_{nl \rightarrow nl'}^{\rm SSC} = 9.8 \times 10^{-6} \times n^4 
    \quad  [{\rm cm}^3 {\rm s}^{-1}] 
    \label{SSC ev4}
\end{equation}
and
\begin{equation}
     q_{nl\rightarrow nl'}^{\rm BT} = 1.23 \times 10^{-5} \times n^4 
     \quad [{\rm cm}^3 {\rm s}^{-1}] , 
     \label{BT ev last}
\end{equation}

We can see the discrepancies more clearly if we take the limit of large-$l$ for both the BTM and SSC expressions given by eq. \eqref{BTMrate} and eq. \eqref{SSCeq_1} (for $Z_p=Z_t=1$ and $T=100$~K):

\be
    q_{nl\rightarrow nl'}^{\rm BTM}\sim 9.80\times10^{-5} \times l_>^2 \quad  [{\rm cm}^3 {\rm s}^{-1}] 
    \label{BT ev lrg3}
\ee
and
\be
      q_{nl \rightarrow nl'}^{\rm SSC} \sim  3.92\times10^{-5} l_{>}^3 \quad [{\rm cm}^3 {\rm s}^{-1}] \quad . 
      \label{SSC ev lrg3}
\ee

Thus

\be
q_{nl \rightarrow nl'}^{\rm SSC}/
q_{nl\rightarrow nl'}^{\rm BTM}
\sim 0.4 l_> \quad\quad \mbox{as}\quad\quad l\rightarrow n\,.
\ee

A full derivation of the analytic evaluation for both intermediate and large $l$-changing collisions is given in appendix~\ref{analytic evaluation}.

The analytical comparison in each transition range has quantified the numerical discrepancies in the rate coefficients. 
We obtained additional numerical results, using the {\sc autostructure} code, including an imposed restriction on the contribution of higher multipoles in the Born approximation. Specifically, we progressively limit the number of contributing multipoles in the summation of Eq. \eqref{Borncross}, first considering only the lowest-order term, then successively adding higher-order terms. The summation starts at \( \lambda = 2 \) and increases in steps of \( \Delta \lambda = 2 \), due to parity conservation.

\begin{figure*}
    \includegraphics[width=\textwidth]{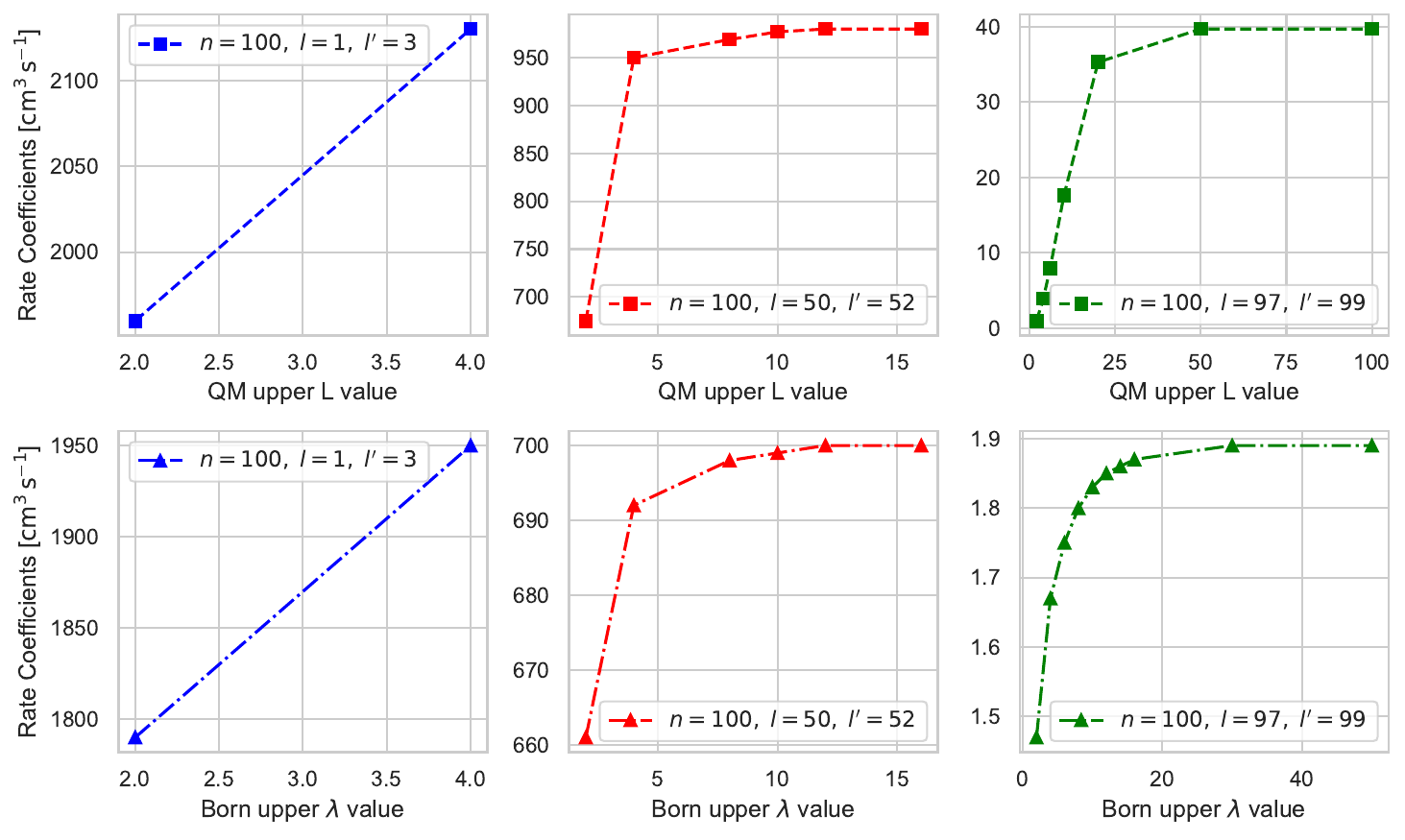}
    \caption{Variation of Born and TDSE QM rate coefficients [cm$^3$ s$^{-1}$] with the restriction of multipoles, at T=100 [K]. }
    \label{Multipole_Restriction}
\end{figure*}

Figure \ref{Multipole_Restriction} clearly shows an increasing relative contribution from the higher pseudo-multipole terms to the TDSE QM $L$-sum compared to the Born multipole sum; for $l \gtrsim n/2$ the two summations disagree increasingly. It is evident that the first multipoles significantly influence and dominate the Born approximation's summation term, with the remaining multipoles contributing only $10-30 \%$. This raises the question of whether the disagreement between the Born and the TDSE QM approximations is only observed in quadrupole $l$-changing collisions and persists for higher multipole transitions. Comparison with TDSE QM approximation is essential for conclusive insights, especially given the BT Bethe approximation's internal agreement with the Born approximation. The agreement between the BT and the Born approximation would be worse if higher multipoles were restricted.

Referring to the Born summation term, eq. \eqref{matrix element 1},
BT obtained the quadrupole collision strength without the contribution of higher multipoles. For higher $l$ values, the multipoles physically contribute more to the sum as $l \rightarrow n$, compared to the first term of the sum. Considering the efficacy of the Born approximation in small $l$-changing collisions, any observed disagreement likely stems from the influence of higher multipoles on the summation term. We therefore suggest that the contribution of higher multipoles plays a pivotal role in the observed disagreement between the Born and the TDSE QM approximations, especially for large $l$-changing collisions. The quadrupole Born terms always dominate the total expression when summing over all multipoles without a limit. This completely differs from how the TDSE QM approximations vary based on its summation term. The Born approximation's multipole expansion involves distinct electric multipole interactions (e.g., dipole, quadrupole, octupole) integral to the electrostatic interaction governing the collision process.

We applied a limiting parameter to the TDSE QM approximation's summation term, eq. \eqref{VR01 QM Prob}, further investigating the origin of the discrepancy with the Born. The limiting parameter whose impact is illustrated in Fig. \ref{Multipole_Restriction}, is the only parameter that can restrict the contribution of higher multipoles on the expansion. We truncate the sum at that $L$ value,  beyond which the rate coefficient is constant with no further fluctuations. The Born and the TDSE QM behave in the same way inasmuch as they yield consistent results when higher multipoles are unrestricted.

We observed that the Bethe approximation, the first term in the Born summation \citep{Burgess2005}, aligns well with the TDSE QM for small and intermediate $l$ transitions, but deviates for large quadrupole $l$-changing collisions. We also noticed that when restricting the summation term to the first quadrupole multipole, the TDSE QM agrees approximately with the Born rate coefficient, including all multipoles. This further agreement between the Born and the TDSE QM approximations supports that the primary source of disagreement between the latter for large $l$-changing collisions arises from the contribution of higher multipoles to their respective summation terms.

We find mathematical disagreement between the summation terms of the Born \eqref{matrix element 1} and the TDSE QM \eqref{VR01 QM Prob} approximations when higher multipoles start to contribute. The Born summation follows a pattern based on $|l-l'|$ and extends up to $l+l'$ in an alternating sequence, while there is no corresponding physical interpretation of the TDSE QM sum.

\section{Tests on the relevance of quadrupole rates}
\label{Applications}

\begin{table}
    \centering
    \caption{Comparison table of dipole and quadrupole quantum mechanical (QM) rate coefficients [cm$^3$ s$^{-1}$] for transitions starting at the same $l$ for the same $n$, at T=100 [K] and N$_e$=1 [cm$^3$].}
    \begin{tabular}{ c | c c | c c }
        \hline
        $n$ & $l \rightarrow l'$ & Dipole & $l \rightarrow l'$ & Quadrupole \\ 
        \hline
        10 & 1-2 & 1.08 $\times 10^{1}$ & 1-3 & 2.04  $\times 10^{-1}$\\
        10 & 4-5 & 7.11 $\times 10^{0}$ & 4-6 & 1.15  $\times 10^{-1}$\\
        10 & 7-8 & 3.39  $\times 10^{0}$ & 7-9 & 4.54  $\times 10^{-2}$ \\
        30 & 1-2 & 7.20 $\times 10^{2}$ & 1-3 & 1.72 $\times 10^{1}$ \\
        30 & 14-15 & 4.27 $\times 10^{2}$ & 14-16 & 8.33 $\times 10^{0}$ \\
        30 & 27-28 & 7.97 $\times 10^{1}$ & 27-29 & 1.10 $\times 10^{0}$\\ 
        50 & 1-2 & 4.90 $\times 10^{3}$& 1-3 & 1.33  $\times 10^{2}$ \\
        50 & 25-26 & 2.79 $\times 10^{3}$ & 25-27 & 6.11  $\times 10^{1}$ \\
        50 & 47-48 & 3.41 $\times 10^{2}$ & 47-49 & 5.02  $\times 10^{0}$ \\
        100 & 1-2 & 6.40 $\times 10^{4}$ & 1-3 & 2.13 $\times 10^{3}$ \\
        100 & 20-21 & 4.71 $\times 10^{4}$ & 20-22 & 1.31 $\times 10^{3}$ \\
        100 & 50-51 & 3.76 $\times 10^{4}$ & 50-52 & 9.80 $\times 10^{2}$ \\
        100 & 97-98 & 2.42 $\times 10^{3}$ & 97-99 & 3.97 $\times 10^{1}$ \\
        250 & 1-2 & 1.75 $\times 10^{6}$ & 1-3 & 8.34 $\times 10^{4}$ \\
        250 & 50-51 & 1.27 $\times 10^{6}$ & 50-52 & 4.99 $\times 10^{4}$ \\
        250 & 125-126 & 1.10 $\times 10^{6}$ & 125-127 & 3.83 $\times 10^{4}$ \\
        250 & 247-248 & 3.11 $\times 10^{4}$ & 247-249 & 6.15 $\times 10^{2}$ \\
        500 & 1-2 & 1.90 $\times 10^{7}$ & 1-3 & 1.33 $\times 10^{6}$ \\
        500 & 100-101 & 1.38 $\times 10^{7}$ & 100-102 & 7.91 $\times 10^{5}$ \\
        500 & 250-251 & 1.12 $\times 10^{7}$ & 250-252 & 6.31 $\times 10^{5}$ \\
        500 & 497-498 & 2.08 $\times 10^{5}$ & 497-499 & 4.98 $\times 10^{3}$ \\
        \hline
    \end{tabular}
    \label{tab:comp_table}
\end{table}

Regardless of the discrepancies at high $l$, the fundamental question
for astrophysical applications is whether quadrupolar rates affect the level populations of spectroscopically important states, and should be included in collisional-radiative models, alongside the 
dipolar rates. 

To start with, we have compared the dipolar and quadrupolar rate coefficients for transitions starting at the same $l$ for the same $n$. A range of values is shown in Table~\ref{tab:comp_table}.
It is clear that the quadrupolar rates are typically 14--100 times weaker than the 
dipolar ones. 

Considering the limiting cases for both rates, we find that they depend similarly on $n$ and the nuclear charge, $Z_t$ Hence, we expect that also for any other H-like atom or ion the quadrupole rates are negligible compared to the dipolar rates. 

Still, one could wonder how much the inclusion of the quadrupolar rate 
changes the populations of all the states.  To answer this question, the hydrogenic collisional radiative model, described in \cite{1987MNRAS.224..801H} and based on \cite{1971MNRAS.153..471B}, was modified to include the BTM rates, chosen as an upper limit for the low-$l$ states which prove to be most affected.

Departure coefficients and line emissivities were compared with and without quadrupole transitions for $10^2\le \rm T [K] \le 10^6$, $10^2 \le \rm N_e [\rm cm^{-3}] \le 10^{12}$ and $2 \le n \le 100$ in Case A and Case B, depending on whether the Lyman lines were assumed optically thin or thick respectively. The largest effects were found at the lowest temperatures and densities for Case A, where departure coefficients changed by up to 8\%, while emissivities changed by 5\%. At typical nebular temperatures of 10$^4$ K no departure coefficients or emissivity was changed by more than 1\%. We note that a direct comparison of the quadrupole $l \to l+2$ with the dipole $l \to l+1$ rate coefficient does not capture all the physics, since the probability of population being transferred from a state $n l$ to a state $n l+2$ via the dipole $l \to l+1 \to l+2$ collisional process must also take into account the probability of a radiative decay from the $n l+1$ state, which does not affect the quadrupole $l \to l+2$ process.

In general, the impact of quadrupole transitions is expected to be greatest for those $n$ above which radiative processes dominate the populations and below which populations are statistically distributed by rapid $l$-changing collisions. Thus the contribution of quadrupole rates is less significant at $n$ values above and below this $n$ regime, whose position is a function of the ambient electron density. For example \cite{Guzman2019}, state that the critical densities, at which radiative and collisional processes are in balance, are $30$ cm$^{-3}$ and $0.6$ cm$^{-3}$, for $n=30$ and $n=50$  respectively. In Table~\ref{tab:qeffect_table} we show the transitions which exhibit the largest change when quadrupole transitions are added, as a function of electron temperature and density,  for a range of conditions, including those typically found in photoionized nebulae.

\begin{table}
    \centering
    \caption{Table showing the maximum percentage change in emissivity due to the inclusion of quadrupole collision rates, at a selection of temperatures [K] and electron densities [cm$^3$], and the upper, $n_u$, and lower $n_l$, principal quantum numbers of the transition for which it occurs. All transitions with $n_u \le 100$ were searched. }
    \begin{tabular}{ l | c c | l }
        \hline
        T[K] & Electron & Transition & maximum \%age  \\
              & density [cm$^3$] & $n_u - n_l$ & emissivity change \\
        \hline
        10$^2$ & 10$^2$ & 35-2 & 6.0 \\
            &   10$^4$ & 19-2 & 4.8 \\
            &  10$^6$ & 12-11 & 3.7 \\
            & 10$^8$ & 6-5 & 3.3 \\   
            & 10$^{10}$ & 5-4 & 2.3 \\   
     10$^{2.5}$ & 10$^2$ & 38-2 & 4.3 \\
            &   10$^4$ & 20-2 & 3.7 \\
            &  10$^6$ & 12-11 & 2.8 \\
            & 10$^8$ & 6-5 & 2.3 \\   
                  & 10$^{10}$ & 4-3 & 0.88 \\   
   10$^3$ & 10$^2$ & 43-2 & 2.9 \\
            &   10$^4$ & 22-2 & 2.6 \\
            &  10$^6$ & 13-12 & 2.0 \\
            & 10$^8$ & 7-6 & 1.7 \\   
            & 10$^{10}$ & 4-3 & 0.66 \\   
         10$^{3.5}$ & 10$^2$ & 47-2 & 1.8 \\
            &   10$^4$ & 23-2 & 1.7 \\
            &  10$^4$ & 14-13 & 1.4 \\
            & 10$^4$ & 7-6 & 1.3 \\   
            & 10$^{10}$ & 4-3 & 0.51 \\   
          10$^4$ & 10$^2$ & 53-2 & 1.0 \\
            &   10$^4$ & 28-2 & 0.93 \\
            &  10$^6$ & 14-13 & 0.90 \\
            & 10$^8$ & 7-6 & 0.84 \\   
            & 10$^{10}$ & 4-3 & 0.35 \\   
        10$^{4.5}$ & 10$^2$ & 52-2 & 0.52 \\
            &   10$^4$ & 29-2 & 0.47 \\
            &  10$^6$ & 16-14 & 0.48 \\
            & 10$^8$ & 8-7 & 0.44 \\   
            & 10$^{10}$ & 5-4 & 0.21 \\   
   \hline
    \end{tabular}
    \label{tab:qeffect_table}
\end{table}

\section{Conclusions}
\label{sec:concludes}

We have developed a simple analytic formula (BTM) for quadrupole $l$-changing collisions which will be of interest to  model astrophysical plasma emission, 
for example the hydrogen and helium-like recombination spectra \citep{Badnell2021, Guzman16}.

We find a good agreement between this BTM formula and results from our quantal Born calculations
using the {\sc autostructure} code, as well as between the SSC formula and the TDSE QM approximation. 

We also find good agreement between our Born results and those we obtain from the TDSE method \citep{Vrinceanu2012} for small-$l$ $(<< n)$, but the two start to differ for intermediate-$l$ $(\sim n/2)$ and diverge increasingly
for large-$l$ $(\gtrsim n/2)$. 

The source of the disagreement between the two quantal results appears to
lie with their representation of the higher ($>2$) terms of the multipole expansion, which is much larger in the quadrupole case
than the dipole one. 
There is a much larger relative contribution from the higher pseudo-multipole terms $(L=2,3,4,\ldots n/2)$
on the TDSE QM total, compared with the contribution of the higher multipoles ($\lambda = 2, 4, 6,\ldots l+l'$) to the Born one. Higher multipoles contribute significantly (approximately 90\%) to the TDSE QM approximation, while applying limiting parameters on multipole expansion terms restricts their contribution and allows for a closer examination of rate coefficient variations. At this evaluation level, discrepancies between the TDSE and Born approximations highlight differences in how quadrupole $l$-changing collisions are treated. Given that TDSE is a full quantum approach, it is expected to provide a more rigorous description. However, further investigation is needed to fully understand the physical implications of these differences.

However, regardless of the reason for the discrepancy, it is clear that the quadrupolar rates are always going to be significantly lower than the dipolar ones, by typically two orders of magnitude. 
Their inclusion in a hydrogenic collisional radiative model for hydrogen shows minimal changes (a few percent) in the line emissivities, which we consider negligible for most astrophysical applications.

This is the last of a series of papers that began with \cite{Pengelly1964},
advanced to the work of 
\citet{Vrinceanu2001, Vrinceanu2012, Vrinceanu2017, Vrinceanu2019}
leading to our studies
\cite{Guzman16, Guzman17, Guzman2019, Badnell2021}.

ad astra, Nigel

\section{Data Availability}
Results shown in the Figures and Tables will be shared on reasonable request to the corresponding author.

\section{Acknowledgements}
This manuscript is dedicated to the memory of Nigel Badnell. He contributed to large parts of this paper and worked for a long time on this topic. He 
received support from STFC (UK) through the University of Strathclyde APAP Network grant ST/R000743/1. ED worked on this topic with Nigel Badnell as an undergraduate student, and now  acknowledges support from STFC (UK) via a PhD studentship. 
GDZ acknowledges support from STFC (UK) via the consolidated grants to the atomic astrophysics group at DAMTP, University of Cambridge (ST/P000665/1. and ST/T000481/1).

\bibliographystyle{mnras}
\bibliography{paper} 




\appendix

\section{Born Approximation by NRB}
\label{append_born}

Consider a transition $q-q'$ between atomic states of hydrogen, where $q=n l m$, ignoring spin. Then, 

\be
Q(q\rightarrow q')= \frac{K_{q'}}{K_{q}} \int |F_{q'q}(\mathbf{K}_q, \mathbf{K}_{q'})|^2 \; \mathrm{d}\Omega \;,
\label{born_1}
\ee

where $\mathbf{K}_q$ is the initial momentum of the free electron.  In the Born approximation the scattering amplitude is given by

\be
F^{\rm B}_{q'q}(\mathbf{K}_q, \mathbf{K}_{q'}) = \frac{1}{2\pi} \int \phi_{\mathbf{K}_{q'}} (\mathbf{r_2}) V_{q'q}(\mathbf{r_2})\, \phi_{\mathbf{K}_q} (\mathbf{r_2}) \; \mathrm{d}\mathbf{r_2} \;,
\label{Born_3}
\ee

where $V_{q'q}$ is given by 

\be
V_{q'q} (\mathbf{r_2}) = \int \phi^*_{q'}(\mathbf{r_1}) \left( \frac{1}{r_{12}}-\frac{1}{r_2}\right) \phi_q(\mathbf{r_1}) \; \mathrm{d}\mathbf{r_1} \;,
\label{Born_4}
\ee

where, 

\be
\phi_{\mathbf{K}_q}(\mathbf{r})=e^{i\mathbf{K}_{q} \cdot \mathbf{r}} \;
\label{Born_5}
\ee

is a plane wave solution of 

\be
(\nabla^2+K_{q}^2) \, \phi_{\mathbf{K}_q}=0 \;
\label{Born_6}
\ee

and $\phi_q(r)$ are atomic wavefunctions of the form 

\be
\phi_q(\mathbf{r})=Y_{lm}(\mathbf{\hat{r}}) \frac{P_{nl}(r)}{r} \;,
\label{Born_7}
\ee

where $Y_{lm}(\mathbf{\hat{r}})$ is a spherical harmonic and 

\be
\frac{1}{r_{12}} = \frac{1}{|\mathbf{r}_1 - \mathbf{r}_2|} \;.
\label{Born_8}
\ee

To see a formal (partial wave) solution with application to Coulomb Born, see \cite{BHT_1970}, pp 226-230. 

To proceed with the Born solution we do not need to make a partial wave analysis. Substituting eq.~\eqref{Born_4} into eq.~\eqref{Born_3}, we obtain, 

\begin{multline}
F^{B}_{q'q}(\mathbf{K}_q, \mathbf{K}_{q'}) = -\frac{1}{2\pi} \int \phi^*_{q'}(\mathbf{r}_1) \phi_q(\mathbf{r}_1) \\
\times \int \left(\frac{1}{r_{12}} - \frac{1}{r_2}\right) e^{i\mathbf{K} \cdot \mathbf{r}_2} \, \mathrm{d}\mathbf{r}_2 \, \mathrm{d}\mathbf{r}_1 \;,
\label{Born_9}
\end{multline}

where $\mathbf{K}=\mathbf{K}_q - \mathbf{K}_{q'}$ is the momentum transfer

A standard result in the literature comes from \cite{Bethe_1930} integral, given by 

\be
\int \frac{e^{i \mathbf{K} \cdot \mathbf{r}_2}}{|\mathbf{r}_1-\mathbf{r}_2|} \mathrm{d}\mathbf{r}_2 = \frac{4\pi}{K^2} e^{i \mathbf{K} \cdot \mathbf{r}_1} \quad \mbox{and} \quad 
\int \frac{e^{i \mathbf{K} \cdot \mathbf{r}_2}}{r_2} \mathrm{d}\mathbf{r}_2 = \frac{4\pi}{K^2} \; . 
\label{Born_11}
\ee

Also, we always use orthonormal atomic wavefunctions 

\be
\int \phi^*_{q'} (\mathbf{r}_1) \phi_q (\mathbf{r}_1) \mathrm{d}\mathbf{r}_1 = \delta_{qq'} \; .
\label{Born_12}
\ee

Using eq.~\eqref{Born_11} and eq.~\eqref{Born_12} in eq.~\eqref{Born_9}, we obtain, 

\be
F^{B}_{q'q}(\mathbf{K}) = -\frac{2}{K^2} \left[\int e^{i \mathbf{K} \cdot \mathbf{r}} \phi^*_{q'}(\mathbf{r}_1) \phi_q(\mathbf{r}_1) \mathrm{d}\mathbf{r}_1 - \delta_{qq'} \right] \; ,
\label{Born_14}
\ee

where,

\be
[...] = <q'|e^{i \mathbf{K} \cdot \mathbf{r}_1} -1 |q> .
\label{Born_15}
\ee

We can re-write the solid angle as 

\be
\mathrm{d}\Omega = \mathrm{d}\phi \sin{\theta} d\theta \; .
\label{Born_16}
\ee

Since we orientate $K_q$ along the z-axis, then

\be
K^2= \mathbf{K} \cdot \mathbf{K} = (\mathbf{K}_{q'} - \mathbf{K}_{q}) \cdot (\mathbf{K}_{q'} - \mathbf{K}_{q}) = K_{q'}^2 + K_q^2 - 2 K_q K_{q'} \cos{\theta} \;.
\label{Born_17}
\ee

So, 

\be
2K\, \mathrm{d}K = 2\, K_q\, K_{q'} \sin{\theta} \, \mathrm{d}\theta
\label{Born_18}
\ee

and 

\be
K_q\, K_{q'} \int \sin{\theta} \, \mathrm{d}\theta = \int K \; \mathrm{d}K \; .
\label{Born_19}
\ee

Hence, using eq.~\eqref{Born_14} and eq.~\eqref{Born_19} into eq.~\eqref{born_1}, we obtain, 

\be
Q_{(q \rightarrow q')} = \frac{\pi}{K_q^2} \int \frac{\mathrm{d}\phi}{2\pi} \int \frac{8 \mathrm{d}K}{K^3} |<q'|e^{i \mathbf{K} \cdot \mathbf{r}_1} -1 |q>|^2\; .
\label{Born_20}
\ee

The collision strength $\Omega_{q'q}$ is defined from, 

\be 
Q_{q'q} = Q(nlm \rightarrow n'l'm') = \frac{\pi}{K_q^2} \, \Omega_{q'q}\; .
\label{Born_22}
\ee

In order to match results obtained by \cite{Burgess1997}, expand the plane wave in spherical harmonics  \citep{edmonds1996angular}, then 

\be
e^{i \mathbf{K} \cdot \mathbf{r}} = 4\pi \sum_{\lambda \mu} i^{\lambda} \, j_\lambda \, (Kr) \, Y_{\lambda \mu}^* \, (\mathbf{\hat{K}}) \, Y_{\lambda \mu} \, (\mathbf{\hat{r}})\; .
\label{Born_23}
\ee

Then, 

\be \Omega_{q'q} = 8 \int \frac{\mathrm{d}K}{K^3} \; |E_{q'q}(K) |^2\; , 
\ee
where
\be
E_{q'q}(K) = 4\pi \sum_{\lambda \mu} i^{\lambda} \, Y_{\lambda \mu}^* \, (\mathbf{\hat{K}}) <q' |  B_{\lambda \mu} | q>\; 
\ee
and 
\be
B_{\lambda \mu} = Y_{\lambda \mu}(\mathbf{\hat{r}}) (j_\lambda(Kr) - \delta_{\lambda_0})\; .
\ee
Then
\be
|E_{q'q}(K)|^2 = 4\pi \sum_{\lambda \mu} | <q'|B_{\lambda \mu}|q>|^2,
\label{Born_24}
\ee

which still includes the magnetic quantum numbers m, m'.

The Wigner-Eckart theorem \citep{edmonds1996angular} enables us to factor our the magnetic quantum numbers. Hence, 

\begin{align}
<n' l'm'| B_{\lambda \mu} | n l  m > &= (-1)^{l' - m'} \begin{pmatrix} l' & \lambda & l \\ -m' & \mu & m \end{pmatrix} \nonumber \\
&\quad \times <n' l' \parallel \mathbf{B_{\lambda}} \parallel n l>
\label{Born_26}
\end{align}

and reduces the $B_{\lambda \mu}$ dependence to $\mathbf{B_{\lambda}}$, as $\mu$ is the "magnetic" component of $Y_{\lambda \mu}$. Then the reduced matrix element is  

\begin{align}
<n' l' \parallel \mathbf{B_{\lambda}} \parallel n l > &= (-1)^{l'} \left[\frac{(2l +1)(2\lambda+1)(2l' +1)}{4\pi}  \right]^{1/2} \nonumber \\
&\quad \times \begin{pmatrix} l' & \lambda & l \\ 0 & 0 & 0 \end{pmatrix} <l' \parallel \mathbf{j_\lambda} \parallel l>\; ,
\label{Born_27}
\end{align}

where, 

\be 
<l' \parallel \mathbf{j_\lambda} \parallel l> = \int_0^{\infty} \mathrm{d}r\, P_{n' l'} (r) [j_{\lambda} (Kr) - \delta_{\lambda 0}] P_{n l} (r)\; .
\label{Born_28}
\ee

The Wigner 3-j symbol (Edmonds 1957, p.125) is given by, 

\begin{align}
\begin{pmatrix} l' & \lambda & l \\ 0 & 0 & 0 \end{pmatrix} &= (-1)^{L/2} \left[ \frac{(l' + \lambda - l)(l' + l - \lambda)!}{(l' + \lambda + l + 1)!} \right]^{1/2} \nonumber \\
&\quad \times \left[ \frac{(L/2)!}{(L/2 - l')!(L/2 - \lambda)!(L/2 - l)!} \right]^{1/2}\; ,
\label{Born_29}
\end{align}

where, $L = l' + \lambda + l$

The magnetic quantum numbers are then easily eliminated as in \cite{Burgess2005}, given also that, 

\be 
\sum_{m' m \mu} \begin{pmatrix} l' & \lambda & l \\ m & \mu & m' \end{pmatrix}^2 = 1\; .
\label{Born_30}
\ee 

Which finally leaves us with equations \eqref{Borncross} and \eqref{Born Collision strength} as presented in the paper. 

\section{BTM Approximation by NRB}
\label{append_BTM}

In this appendix we describe the modifications made to the \cite{Burgess2005} approximation, which we now called BTM (Burgess and Tully Modified), especially for the analytic form of the quadrupole line strength of the target ($S_2$), given in eq.~\eqref{line2} in the paper. 

Looking at eq. 17, eq.31 and eq. 33 from the \cite{Burgess2005} that present the collision strength, there had to be modifications to the quadrupole line strength of the target term, $S_2$. Also, the rate coefficients are given in terms of energy, however zero excitation energy is assumed for quadrupole $l$-changing collisions. 

\cite{Burgess2005} present the collision strength in their equation 33 as, 

\be
\Omega_{ij} = \left(\frac{2}{15}\right) M^2 \, Z_p^2 \, S_2 \, \left(\frac{\alpha}{r_0} \right)^2\; ,
\label{BTM_1}
\ee

where, $\alpha^2=6$ agrees with eq.17 from the \cite{Burgess2005} paper and $r_0$ can be determined by infinite energy or by $r_0=\frac{c_0 n^2}{Z_t}$

The quadrupole line strength for the target transition, $S_2$ is given by, 

\begin{align}
S_2(nl', nl'=l \pm 2) &= (2)(2l+1)(2l'+1) \begin{pmatrix} l & 2 & l' \\ 0 & 0 & 0 \end{pmatrix} \nonumber \\
&\quad \times \left| \int P_{nl} \, r^2 \, P_{nl'}\, \mathrm{d}r\, \right|^2\; ,
\label{BTM_S2}
\end{align}

where, 

\be
(2)(2l+1)(2l'+1) \begin{pmatrix} l & 2 & l' \\ 0 & 0 & 0 \end{pmatrix}^2 = \frac{3 l_> (l_>-1)}{(2l_>-1)}
\label{BTM_S2_2}
\ee

and 

\be
\left| \int P_{nl} \, r^2 \, P_{nl'}\, \mathrm{d}r\, \right|^2 = 100 \left(\frac{n a_a}{2Z_t} \right)^2 (n^2-l_>^2)^2 (n^2-(l_>-1)^2)\; .
\label{BTM_S2_3}
\ee

Hence, we obtain the quadrupole line strength of the target term, $S_2$, as we presented in eq.~\eqref{line2} in the paper, which is, 

\be
S_2=\left( \frac{300}{2} \right) \left( \frac{n}{2 Z_t} \right)^4 \left (\frac{l_>(l_>-1)}{(2l_>-1)} \right)
      (n^2-l_>^{2})(n^2-(l_>-1)^2)\; ,
\label{line2_appendix}
\ee 

Assuming that $l>>1$, and given that $l'=l \pm 2$ then eq.\eqref{line2_appendix} simplifies to

\be
S_2 = \frac{300}{2} \left(\frac{n}{2Z_t}\right)^4 l (n^2-l^2)^2\; .
\label{line2_appendix_2}
\ee

\section{Analytic Evaluation}
\label{analytic evaluation}

The analytic rate coefficient of BTM is given by substituting the collision strength, quadrupole line strength and the effective size of the atom, as follows,

\begin{equation}
\begin{split}
     q_{nl\rightarrow nl'}^{BT} & =\frac{2\sqrt{\pi M}}{\sqrt{kT(Ryd)}} \left( \frac{12}{15} \right) \left( \frac{Z_t}{c_0 \, n^2} \right)^2 \left( \frac{300}{2} \right) \left( \frac{n^4}{16 Z_t^{4}} \right) \\
     & \left (\frac{l_>(l_>-1)}{(2l_>-1)} \right) \times (n^2-l_>^{2})(n^2-(l_>-1)^2)\times \frac{1}{(2l+1)} \frac{a_0^3}{\tau_0}\; ,
     \label{BT ev1}
\end{split}
\end{equation}

which is expressed in atomic units while the temperature (T) is expressed in Rydberg units. \cite{Burgess2005} use

\be
2\pi^{1/2} \left(\frac{a_0^3}{\tau_0}\right)=2.17167 \times 10^{-8} \quad [\rm cm^3 \rm s^{-1}]\; ,
\ee

to convert atomic units to cgs units, which we follow as the SSC rate coefficient expression is given in cgs units. By making use of the appropriate values for constants, and for T=100 K, we simplify eq.~\eqref{BT ev1} as a numerical part, times an expression only written in terms of $n$ and $l$, given by 

\begin{equation}
\begin{split}
    q_{nl\rightarrow nl'}^{\rm BT} = 4.90\times10^{-5} \times \left[\frac{1}{2l+1} \times \frac{l_>(l_>-1)}{2l_>-1} \right. \\
    \left. \times (n^2 - l_>^2) (n^2 - (l_> - 1)^2) \right] \quad [\rm cm^3 \rm s^{-1}]\; .
\end{split}
\label{BT 2nd RC}
\end{equation}

We first consider approximate expressions for the case of intermediate $l$ values, such that $1<<l<<n$. For $l>>1$ we then have,  

\begin{equation}
     q_{nl\rightarrow nl'}^{\rm BT}=4.90\times10^{-5} \frac{l_> (n^2-l_>^2)^2 }{2(2l_>+1)} \quad [\rm cm^3 \rm s^{-1}]\; ,
     \label{BT ev2}
\end{equation}

which further simplifies to  

\begin{equation}
    q_{nl\rightarrow nl'}^{\rm BT} = 1.23 \times 10^{-5} \; \times (n^2 -l^2)^2 \quad [\rm cm^3 \rm s^{-1}]\; .
    \label{BT ev3}
\end{equation}

Now, the full SSC simple rate coefficient, from \cite{Vrinceanu2012} is given by, 

\begin{equation}
\begin{split}
    q_{nl\rightarrow{}nl'}^{SSC} = 1.294 \times 10^{-5} \sqrt{\frac{M}{m_{e}}} \frac{Z_p^2}{\sqrt{T}} \times \\
    \frac{n^2[n^2(l+l')-l_{<}^2(l+l'+2|\Delta l|)]}{(l+1/2)|\Delta l|^3} \quad [\rm cm^3 \rm s^{-1}]\; .
\end{split}
\label{SSCeq_2}
\end{equation}

Then, we can multiply eq.~\ref{SSCeq_1} by a factor fo $2/2$ so that the denominator can be easily aligned in a way that it matches the initial statistical weight of the BT rate coefficient of the form $1/(2l+1)$. We then obtain,

\begin{equation}
\begin{split}
    q_{nl \rightarrow nl'}^{SSC} = 3.9213 \times 10^{-5} \times \\
    \frac{n^2[n^2(l_>-1)-(l_>-2)^2(l_>+1)]}{2(2l+1)} \quad [\rm cm^3 \rm s^{-1}]\; .
\end{split}
\label{SSC ev1}
\end{equation}

Then, similarly with the BTM evaluation, we can take the limit where $l>>1$ and simplify the SSC rate coefficient as, 

\begin{equation}
    q_{nl \rightarrow nl'}^{SSC} = 9.8 \times 10^{-6} \times n^2 \left[n^2-l^2\right] \quad [\rm cm^3 \rm s^{-1}]\; .
    \label{SSC ev2}
\end{equation}

If in addition $l<<n$, then the expression further simplifies to determine the final results of the analytic comparison for intermediate $l$-changing transitions.

\begin{equation}
    q_{nl \rightarrow nl'}^{\rm SSC} = 9.8 \times 10^{-6} \times n^4 \quad [\rm cm^3 \rm s^{-1}]
    \label{SSC ev4_apend}
\end{equation}
and
\begin{equation}
     q_{nl\rightarrow nl'}^{\rm BT} = 1.23 \times 10^{-5} \times n^4 \quad [\rm cm^3 \rm s^{-1}]\; .
     \label{BT ev last_apend}
\end{equation}

The SSC and the BTM expressions are finally compared for large $l$-changing transitions where the biggest disagreement is observed both numerically and analytically. The assumptions that hold for large $l$-changing transitions are slightly different from intermediate $l$-changing transitions. For large transitions, the SSC is in a highly accurate agreement with the TDSE QM. Additionally, numerical results show the highest level of agreement between the BTM and the Born. If we take $n = l_> +1 $, the initial SSC approximation,  eq.~\eqref{SSCeq_1}, for the large $l$-changing transitions is simplified into Eq.\eqref{SSC ev lrg2}.

\begin{equation}
\begin{split}
    q_{nl \rightarrow nl'}^{SSC} = & \, 3.92 \times 10^{-5} \times \frac{2}{2} \times \\
    & \, \frac{n^2[2n^2(l_>-1)-2(l_>-2)^2(l_>+1)]}{8(l_>+1/2)} \quad [\rm cm^3 \rm s^{-1}] \\
    = & \, 3.92 \times 10^{-5} \times \frac{1}{2} \times \\
    & \, (l_> + 1)^3 \times \frac{[(l_>^2-1) - (l_>-2)^2]}{(2l_> + 1)} \quad [\rm cm^3 \rm s^{-1}]\; .
\end{split}
\label{SSC ev lrg2}
\end{equation}

The same assumption holds for the BT rate coefficient expression evaluated on large $l$-changing transitions, so Eq.\eqref{BT 2nd RC} is further simplified into Eq.\eqref{BT ev lrg2}.

\begin{equation}
     q_{nl\rightarrow nl'}^{\rm BT}=4.90\times10^{-5} \times \frac{4l_>^2(2l_>+1)(l_>-1)}{(2l+1)(2l_>-1)} \quad [\rm cm^3 \rm s^{-1}]\; .
     \label{BT ev lrg2}
\end{equation}

Also, for large $l$ transitions, the assumption that $l>>1$ still holds, therefore, SSC and BT expressions are further simplified into Eq.\eqref{SSC ev lrg3} and Eq.\eqref{BT ev lrg3} respectively, which are the final analytic results for the evaluation of large $l$ transitions.

\begin{equation}
      q_{nl \rightarrow nl'}^{\rm SSC} =  3.92\times10^{-5} \times l_{>}^3 \quad [\rm cm^3 \rm s^{-1}]
      \label{SSC ev lrg3_2}
\end{equation}
and
\begin{equation}
    q_{nl\rightarrow nl'}^{\rm BT}= 9.80\times10^{-5} \times l_>^2 \times \quad [\rm cm^3 \rm s^{-1}]\; .
    \label{BT ev lrg3_2}
\end{equation}

Therefore the final ratio of SSC/BT for large $l$ transitions is approximately $0.4l_>$. 


\bsp	
\label{lastpage}
\end{document}